\documentstyle[aps,prl,epsf,special]{revtex}

\begin{document}
\draft 
{
\title{Fixed-$N$ Superconductivity: The Crossover from the Bulk to the
  Few-Electron Limit}

\author{Fabian Braun and Jan von Delft}
\address{Institut f\"ur Theoretische Festk\"orperphysik, Universit\"at
  Karlsruhe, 76128 Karlsruhe, Germany}
\date{November 23, 1998}%
\maketitle%
\begin{abstract}
  We present a truly canonical theory of superconductivity in ultrasmall
  metallic grains by variationally optimizing fixed-$N$ projected BCS
  wave-functions, which yields the first full description of the {\em entire
    crossover}\/ from the bulk BCS regime (mean level spacing $d$ $\ll$ bulk
  gap $\tilde\Delta$) to the ``fluctuation-dominated'' few-electron regime
  ($d\gg\tilde\Delta$).  A wave-function analysis shows in detail how the BCS
  limit is recovered for $d\ll \tilde \Delta$, and how for $d \gg \tilde
  \Delta$ pairing correlations become delocalized in energy space.  An earlier
  grand-canonical prediction for an observable parity effect in the spectral
  gaps is found to survive the fixed-$N$ projection.
\end{abstract}
}
\narrowtext 
In the early days of BCS theory, its use of essentially grand-canonical (g.c.)
wave-functions was viewed as one of its most innovative, if not perplexing
features: the variational BCS ansatz for the ground state is a superposition
of states with different electron numbers, although BCS \cite{BCS-57}
themselves had emphasized that the true ground state of an isolated
superconductor must be a state of definite electron number.  That this ansatz
was nevertheless rapidly accepted and tremendously successful had two reasons:
Firstly, calculational convenience -- determining the variational parameters
is incomparably much simpler in a g.c.\ framework, where the particle number
is fixed only on the average, than in a canonical one, where a further
projection to fixed electron number is required; and secondly, it becomes
exact in the thermodynamic limit -- fixed-$N$ projections yield corrections to
the BCS ground state energy per electron that are only of order $N^{-1}$, as
shown e.g.\ by Anderson \cite{Anderson-58} and M\"uhlschlegel
\cite{Muehlschlegel-62}.

Recently, however, a more detailed examination of the range of validity of
BCS's g.c.\ treatment has become necessary, in the light of measurements by
Ralph, Black and Tinkham (RBT) \cite{Ralph-97} of the discrete electronic
spectrum of an individual ultrasmall superconducting grain: it had a charging
energy so large ($E_C \gg \tilde \Delta$) that electron number fluctuations
are strongly suppressed, calling for a canonical description, and the number
of electrons $N$ within the Debye frequency cutoff $\omega_D$ from the Fermi
energy $\varepsilon_F$ was only of order $10^2$, hence differences between
canonical and g.c.\ treatments might become important. Moreover, its mean
level spacing $d\propto N^{-1}$ was comparable to the bulk gap $\tilde\Delta$,
hence it lies right in the {\em crossover regime}\/ between the
``fluctuation-dominated'' few-electron regime ($d\gg\tilde\Delta$) and the
bulk BCS regime ($d \ll \tilde\Delta$), which could not be treated
satisfactorily in any of the recent theoretical papers inspired by these
experiments: the results of
\cite{Golubev-94,vonDelft-96,Braun-97,Braun-98,Matveev-97}, including
the predictions of parity effects, were obtained in a g.c.\ framework; and
Mastellone, Falci and Fazio's (MFF) \cite{Mastellone-98} fixed-$N$ exact
numerical diagonalization study, the first detailed analysis of the f.d.\ 
regime, was limited to $N \le 25$.

In this Letter we achieve the {\em first canonical description of the full
  crossover}.  We explicitly project the BCS ansatz to fixed $N$ (for $N \le
600$) before variationally optimizing it, adapting an approach developed by
Dietrich, Mang and Pradal \cite{Dietrich-64} for shell-model nuclei with
pairing interactions to the case of ultrasmall grains.  This projected BCS
(PBCS) approach enables us (i) to significantly improve previous g.c.\ upper
bounds on ground state energies
\cite{Golubev-94,vonDelft-96,Braun-97,Braun-98}; (ii) to check that a previous
grand-canonical prediction \cite{Braun-98} for an {\em observable}\/ parity
effect in the spectral gaps survives the fixed-$N$ projection; (iii) to find
in the crossover regime a remnant of the ``break-down of superconductivity''
found in g.c.\ studies, at which the condensation energy changes from being
extensive to practically intensive; and (iv) to study this change by an {\em
  explicit wave-function analysis,}\/ which shows in detail how the BCS limit
is recovered for $d\ll \tilde \Delta$, and how for $d \gg \tilde \Delta$
pairing correlations become delocalized in energy space.

\paragraph*{The model.---} 
We model the superconducting grain by a reduced BCS-Hamiltonian which has been
used before to describe small superconducting grains
\cite{vonDelft-96,Braun-97,Braun-98,Matveev-97} (it was phenomenologically
successful for $d\le \tilde \Delta$ \cite{Braun-97,Braun-98}, but probably is
unrealistically simple for $d \gg \tilde \Delta$, for which it should rather
be viewed as toy model):
\begin{equation}
  \label{eq:hamiltonian}
  H = \sum_{j=0,\sigma}^{N-1}\varepsilon_j c^\dagger_{j\sigma}c_{j\sigma} 
  - \lambda \, d \sum_{j,j'=0}^{N-1} c^\dagger_{j+}c^\dagger_{j-}c_{j'-} c_{j'+}.
\end{equation}
The $c^\dagger_{j \pm}$ create electrons in free time-reversed
single-particle-in-a-box states $|j,\pm\rangle$, with discrete, uniformly
spaced, degenerate eigenenergies $\varepsilon_j= j d +\varepsilon_0$.  The
interaction scatters only time-reversed pairs of electrons within $\omega_D$
of $\varepsilon_F$.  Its dimensionless strength $\lambda$ is related to the
two material parameters $\tilde\Delta$ and $\omega_D$ via the bulk gap
equation $\sinh1/\lambda = \omega_D/\tilde\Delta$. We chose $\lambda=0.22$,
close to that of Al \cite{Braun-98}.  The level spacing $d$ determines the
number $N=2\omega_D/d$ of levels, taken symmetrically around $\varepsilon_F$,
within the cutoff; electrons outside the cutoff remain unaffected by the
interaction and are thus neglected throughout.

\paragraph*{Projected Variational Method.---} 
We construct variational ground states for $H$ by projecting BCS-type
wave-functions onto a fixed electron number $N=2n_0+b$ \cite{Dietrich-64},
where $n_0$ and $b$ are the number of electron pairs and unpaired electrons
within the cutoff, respectively. Considering $b=0$ first, we take
\begin{eqnarray}
  \label{eq:dmp-wavefunction}
  |0\rangle = C
   \int_0^{2\pi}\!\!\!\! d\phi\,e^{-i\phi n_0} \prod_{j=0}^{N-1} \Big(u_j+
     e^{i\phi} v_j c^\dagger_{j+}c^\dagger_{j-}\Big) |\mbox{Vac}\rangle,
\end{eqnarray}
where $|\mbox{Vac}\rangle$ is the vacuum state. Both $v_j$, the amplitude to
find a pair of electrons in the levels $|j,\pm\rangle$, and $u_j$, the
amplitude for the level's being empty, can be chosen real \cite{Dietrich-64}
and obey $u_j^2+v_j^2=1$. The integral over $\phi$ performs the projection
onto the fixed electron pair number $n_0$, and $C$ is a normalization constant
ensuring $\langle0|0\rangle=1$.

Doing the integral analytically yields a sum over ${2n_0 \choose n_0}$ terms
(all products in (\ref{eq:dmp-wavefunction}) that contain exactly $n_0$
factors of $v_j c^\dagger_{j+}c^\dagger_{j-}$), which is forbiddingly unhandy
for any reasonable $n_0$. Therefore we follow Ref.~\cite{Dietrich-64} and
evaluate all integrals numerically instead. Introducing the following
shorthand for a general {\em projection integral},\/
\begin{eqnarray*}
  R_n^{j_1 \cdots j_N} \equiv  \int_0^{2\pi} {d\phi \over 2 \pi} \,
  e^{-i(n_0-n)\phi} \prod_{j\neq j_1\cdots j_N}
  (u_j^2+e^{i\phi} v_j^2), 
\end{eqnarray*}
the expectation value ${\cal E}_0 = \langle 0 | H | 0 \rangle$ can be
expressed as
\begin{eqnarray}
\nonumber
{\cal E}_0  =  
  \sum_{j} (2 \varepsilon_j- \lambda d ) |v_j|^2
  \frac{R^j_1}{R_0}
  -\lambda d \sum_{j,k} u_j v_j u_k v_k \frac{R^{jk}_1}{R_0}.
\end{eqnarray}
Minimization with respect to the variational parameters $v_j$ leads to a set
of $2n_0$ coupled equations,
\begin{equation}
  \label{eq:variational}
  2(\hat\varepsilon_j+\Lambda_j)u_jv_j = \Delta_j(u_j^2-v_j^2),
\end{equation}
where the quantities $\hat\varepsilon_j$, $\Lambda_j$ and $\Delta_j$ are
defined by
\begin{eqnarray*}
    \hat\varepsilon_j & \equiv & 
(\varepsilon_j-\lambda d/2)\frac{R_1^j}{R_0}, \qquad
  \Delta_j \equiv {\lambda d} \sum_{k}u_{k}v_{k}
  \frac{R_1^{jk}}{R_0}, \\
  \Lambda_j &\equiv &\sum_{k} 
\left(\varepsilon_j-\frac{\lambda d}2 \right)v_{k}^2
  \left[\frac{R_2^{jk}-R_1^{jk}}{R_0}-
    \frac{R_1^{k}}{R_0}\frac{R_1^{j}-R_0^{j}}{R_0}
  \right]\nonumber\\
&& \hspace*{-11pt} -\frac{\lambda d}{2} \sum_{k,\ell} u_{k}v_{k}u_{\ell}v_{\ell}
 \left[\frac{R_2^{jk\ell}-R_1^{jk\ell}}{R_0}-
   \frac{R_1^{k\ell}}{R_0}\frac{R^j_1-R^j_0}{R_0}\right].
\end{eqnarray*}
We obtain an upper bound on the ground state energy and a set of $v_j$'s,
i.e.\ an approximate wave function, by solving these equations numerically. To
this end we use a formula of Ma and Rasmussen \cite{Ma-77} to express any
$R^{j_1\cdots j_N}_n$ in terms of $R_0$ and all $R^j_0$'s, and evaluate the
latter integrals using fast Fourier transform routines.

Next consider states with $b$ unpaired electrons, e.g.\ states with odd number
parity or excited states: Unpaired electrons are ``inert'' because the
particular form of the interaction involves only electron {\em pairs\/}. Thus
the Hilbert subspace with $b$ specific levels occupied by unpaired electrons,
i.e. levels ``blocked'' to pair scattering \cite{Soloviev-61,Braun-97}, is
closed under the action of $H$, allowing us to calculate the energy, say
${\cal E}_b$, of its ground state $|b\rangle$ by the variational method, too.
To minimize the kinetic energy of the unpaired electrons in $|b\rangle$ we
choose the $b$ singly occupied levels, $j\in B$, to be those closest to the
Fermi surface \cite{Braun-98}. Our variational ansatz for $|b\rangle$ then
differs from $|0\rangle$ only in that $\prod_j$ is replaced by
$\big(\prod_{j\in B}c^\dagger_{j+}\big)\prod_{j\not\in B}$. Thus in all
products and sums over $j$ above, the blocked levels are excluded (the $u_j$
and $v_j$ are not defined for $j\in B$) and the total energy ${\cal E}_b$ has
an extra kinetic term $\sum_{j\in B} \varepsilon_j$.

In the limit $d\to0$ at fixed $n_0d$, the PBCS theory reduces to the g.c.\ BCS
theory of Ref.~\cite{Braun-97} (proving that the latter's $N$-fluctuations
become negligible in this limit): The projection integrals can then be
approximated by their saddle point values \cite{Dietrich-64}; since $\phi=0$
at the saddle, the $R$'s used here are all equal, thus $\Lambda_j$ vanishes,
the variational equations decouple and reduce to the BCS gap equation, and the
saddle point condition fixes the {\em mean\/} number of electrons to be
$2n_0+b$.  To check the opposite limit of $d\gg\tilde\Delta$ where $n_0$
becomes small, i.e.\ the f.d. regime, we compared our PBCS results for ${\cal
  E}_0$ and ${\cal E}_1$ with MFF's exact results \cite{Mastellone-98},
finding agreement to within $1\%$ for $n_0 \le 12$.  This shows that
``superconducting fluctuations'' (as pairing correlations are traditionally
called when, as in this regime, the g.c.\ pairing parameter vanishes
\cite{vonDelft-96}) are treated adequately in the PBCS approach.  Because it
works so well for $d \ll \tilde\Delta$ and $d\gg\tilde\Delta$, it seems
reasonable to trust it in the crossover regime $d\sim\tilde\Delta$ too, though
here, lacking any exact results for comparison, we cannot quantify its errors.

\begin{figure}
  \begin{center}
    \vspace{-3mm}
    \leavevmode
    \epsfxsize72mm
    \epsffile{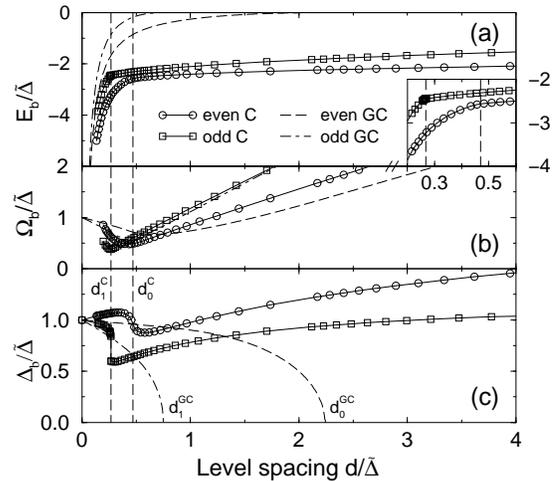}
    \vspace{-2mm}
  \end{center}
  \caption{(a) The ground state correlation energies ${ E}_b$,
    (b) the spectral gaps $\Omega_b = {\cal E}_{b+2} - {\cal E}_b$ and (c) the
    pairing parameters $\Delta_b$, for even and odd systems $(b=0,1)$,
    calculated canonically (C) and grand-canonically (GC) as functions of $d/
    \tilde \Delta = 2 \sinh (1/ \lambda) / N$.  The inset shows a blow-up of
    the region around the characteristic level spacings $d^C_{0} = 0.5 \tilde
    \Delta$ and $d^C_{1} = 0.25 \tilde \Delta$ (indicated by vertical lines in
    all subfigures). The $d^C_b$ (a) mark a change in behavior of ${
      E}_b^{C}$ from $\sim 1/d$ to being almost $d$ independent, and roughly
    coincide with (b) the minima in $\Omega_b$, and (c) the position of the
    abrupt drops in $\Delta_b$.}
  \label{fig:results}
  \label{fig:Omega}
  \label{fig:pairing-parameter}
\end{figure}

\paragraph*{Ground state energies.---}
Figure~\ref{fig:results}(a) shows the ground state condensation energies
$E_b={\cal E}_{b}-\langle F_b|H|F_b\rangle$ for even and odd grains ($b=0$ and
$1$, respectively), which is measured relative to the energy of the respective
uncorrelated Fermi sea ($|F_0 \rangle = \prod_{j<n_0} c^\dagger_{j+}
c^\dagger_{j-} |\text{Vac}\rangle$ or $|F_1\rangle = c^\dagger_{n_0+} |F_0
\rangle$), calculated for $N \le 600$ using both the canonical ($E_{b}^{C}$)
and g.c.\ ($E_{b}^{GC}$) \cite{vonDelft-96,Braun-97} approaches.  The g.c.\ 
curves suggest a ``breakdown of superconductivity''
\cite{vonDelft-96,Braun-97} for large $d$, in that $E_{b}^{GC}= 0$ above some
critical $b$-dependent level spacing $d^{GC}_{b}$.  In contrast, the ${
  E}_b^{C}$'s are (i) significantly lower than the $E^{GC}_b$'s, thus the
projection much improves the variational ansatz; and (ii) negative for {\em
  all}\/ $d$, which shows that the system can {\em always} gain energy by
allowing pairing correlations, even for arbitrarily large $d$. As anticipated
in \cite{Braun-98}, the ``breakdown of superconductivity'' is evidently not as
complete in the canonical as in the g.c.\ case.  Nevertheless, some remnant of
it does survive in $E_{b}^{C}$, since its behaviour, too, changes markedly at
a $b$ (and $\lambda$) dependent characteristic level spacing $d_{b}^C$ ($<
d^{GC}_b$): it marks the end of bulk BCS-like behavior for $d<d_b^C$, where ${
  E}_b^{C}$ is {\em extensive}\/ ($\sim 1/d$), and the start of a f.d.\ 
plateau for $d>d_b^C$, where $E_{b}^{C}$ is practically {\em intensive}
(almost $d$ independent)\cite{likharev}. The
\linebreak
 standard heuristic interpretation
\cite{Tinkham} of the bulk BCS limit $-\tilde\Delta^2/(2d)$ (which is indeed
reached by $E_b^C$ for $d\to 0$) hinges on the scale $\tilde \Delta$: the
number of levels strongly affected by pairing is roughly $\tilde \Delta / d$
(those within $\tilde \Delta$ of $\varepsilon_F$), with an average energy gain
per level of $-\tilde\Delta /2$. To analogously interpret the $d$ {\em
  in}\/dependence of $E_b^{C}$ in the f.d.\ regime, we argue that {\em the
  scale $\tilde \Delta$ loses its significance}\/ -- fluctuations affect {\em
  all}\/ $n_0 = \omega_D/d$ unblocked levels within $\omega_D$ of
$\varepsilon_F$ (this is made more precise below), and the energy gain per
level is proportional to a renormalized coupling $- \tilde \lambda d$
(corresponding to the $1/N$ correction of
\cite{Anderson-58,Muehlschlegel-62} to the g.c.\ BCS result).  
The inset of Fig.~\ref{fig:results}(a) shows the crossover to be quite
non-trivial, being surprisingly abrupt for $E_1^C$.

\paragraph*{Parity Effect.---}
Whereas the ground state energies are not observable by themselves, the
parity-dependent spectral gaps $\Omega_0 = {\cal E}_2-{\cal E}_0$ and
$\Omega_1={\cal E}_{3}-{\cal E}_{1}$ {\em are\/} measurable in RBT's
experiments by applying a magnetic field \cite{Braun-98}.
Figure~\ref{fig:Omega}(b) shows the canonical $(\Omega_b^{C})$ and g.c.\ 
$(\Omega_b^{GC})$ results for the spectral gaps.  The main features of the
g.c.\ predictions are \cite{Braun-98}: (i) the spectral gaps have a minimum,
which (ii) is at a smaller $d$ in the odd than the even case, and (iii)
$\Omega_1<\Omega_0$ for small $d$, which was argued to constitute an
observable parity effect.  Remarkably, {\em the canonical calculation
  reproduces all of these qualitative features, including the parity
  effect}\/, differing from the g.c.\ case only in quantitative details: the
minima are found at smaller $d$, and $\Omega_0^{GC} < \Omega_0^C$ for large
$d$. The latter is due to fluctuations, neglected in ${\cal E}_b^{GC}$, which
are less effective in lowering ${\cal E}_b^C $ the more levels are blocked, so
that $|{\cal E}_b^C - {\cal E}_b^{GC}|$ decreases with $b$.

\begin{figure}
  \begin{center}
    \vspace{-3mm}
    \leavevmode
    \epsfxsize65mm
    \epsffile{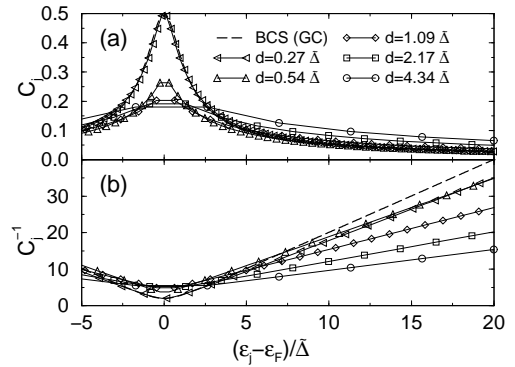}
   \vspace{-2mm}
  \end{center}
  \caption{The pairing amplitudes $C_j$ of Eq.~(\protect\ref{C_j}).
    (a) The dashed line shows the g.c.\ BCS result; pair correlations are
    localized within $\tilde \Delta$ of $\varepsilon_F$.  Lines with symbols
    show the canonical results for several $d$; for $d< d^C_0 \approx
    0.5\tilde\Delta$, the wave functions are similar to the BCS ground state,
    while for $d< d^C_0$ weight is shifted away from $\varepsilon_F$ into the
    tails.  (b) For all $d$, $C_j^{-1} $ shows linear behavior far from
    $\varepsilon_F$.  For larger $d$ the influence of levels far from
    $\varepsilon_F$ increases.}
  \label{fig:state}
\end{figure}

\paragraph*{Wave functions.---} 
Next we analyze the variationally-determined wave-functions. Each $|b \rangle$
can be characterized by a set of correlators
\begin{equation}
  C^2_j (d) =\langle c^\dagger_{j+} c_{j+} c^\dagger_{j-} c_{j-}\rangle
  -\langle c^\dagger_{j+} 
  c_{j+} \rangle\langle c^\dagger_{j-} c_{j-}\rangle, 
  \label{C_j}
\end{equation}
which measure the amplitude enhancement for finding a {\em pair\/} instead of
two uncorrelated electrons in $|j,\pm\rangle$.  For any blocked
single-particle level and for all $j$ of an uncorrelated state one has
$C_j=0$.  For the g.c.\ BCS case $C_j = u_j v_j$ and the $C_j$'s have a
characteristic peak of width $\sim \tilde\Delta$ around $\varepsilon_F$, see
Fig.~\ref{fig:state}(a), implying that pairing correlations are ``localized in
energy space''.  For the BCS regime $d<\tilde\Delta$, the canonical method
produces $C_j$'s virtually identical to the g.c.\ case, {\em vividly
  illustrating why the g.c.\ BCS approximation is so successful: not
  performing the canonical projection hardly affects the parameters $v_j$ if
  $d \ll \tilde\Delta$, but tremendously simplifies their calculation}\/
(since the $2 n_0$ equations in (\ref{eq:dmp-wavefunction}) then decouple).
However, in the f.d.\ regime $d>d^C_{b}$, the character of the wave-function
changes: weight is shifted into the tails far from $\varepsilon_F$ at the
expense of the vicinity of the Fermi energy.  Thus {\em pairing correlations
  become delocalized in energy space} (as also found in \cite{Mastellone-98}),
so that referring to them as mere ``fluctuations'' is quite appropriate.
Fig.~\ref{fig:state}(b) quantifies this delocalization: $C_j$ decreases as
$(A | \varepsilon_j - \varepsilon_F| +B)^{-1}$ far from the Fermi surface,
with $d$-dependent coefficients $A$ and $B$; for the g.c.\ $d=0$ case, $A=2$
and $B=0$; with increasing $d$, $A$ decreases and $B$ increases, implying
smaller $C_j$'s close to $\varepsilon_F$ but a {\em slower fall-off far from
  $\varepsilon_F$.}\/ In the extreme case $d \gg d^C_b$ pair-mixing is roughly
equal for all interacting levels.

To quantify how the {\em total}\/ amount of pairing correlations, summed over
all states $j$, depends on $d$, Fig.~\ref{fig:pairing-parameter}(c) shows the
{\em pairing parameter} $\Delta_b (d) = \lambda d\sum_j C_j$ proposed by
Ralph~\cite{Braun-98,alternative}, calculated with the canonical
$(\Delta_b^C)$ and grand-canonical $(\Delta_b^{GC})$ approaches.  By
construction, both $\Delta_b^{GC}$ and $\Delta_b^{C}$ reduce to the bulk BCS
order parameter $\tilde\Delta$ as $d \to 0$, when $C_j \to u_j v_j$.
$\Delta_{b}^{GC}$ decreases with increasing $d$ and drops to zero at the same
critical value $d^{GC}_{b}$ at which the energy ${ E}^{GC}_{b}$ vanishes
\cite{Braun-98}, reflecting again the g.c.\ ``breakdown of
superconductivity''. In contrast, $\Delta_{b}^{C}$ is non-monotonic and never
reaches zero; even the slopes of $\Delta_e^{C}$ and $\Delta_e^{GC}$ differ as
$d \to 0$, \cite{Golubev-94,vonDelft-96}, illustrating that the $1/N$
corrections neglected in the g.c.\ approach can significantly change the
asymptotic $d \to 0 $ behavior (this evidently also occurs in Fig.~1.b).
Nevertheless, $\Delta_b^{GC}$ does show a clear remnant of the g.c.\ 
breakdown, by decreasing quite abruptly at the same $d^{C}_b$ at which the
plateau in ${ E}^{C}_{b}$ sets in.  For the odd case this decrease is
surprisingly abrupt, but is found to be smeared out for larger $\lambda$. We
speculate that the abruptness is inversely related to the amount of
fluctuations, which are reduced in the odd case by the blocking of the level
at $\varepsilon_F$, but increased by larger $\lambda$.  $\Delta_{b}^{C}$
increases for large $d$, because of the factor $\lambda d$ in its definition,
combined with the fact that (unlike in the g.c.\ case) the $C_j$ remain
non-zero due to fluctuations.

Our quantitative analysis of the delocalization of pairing correlations is
complimentary to but consistent with that of MFF \cite{Mastellone-98}.
Despite being limited to $n_0 \le 12$, MFF also managed to partially probe the
crossover regime from the f.d.\ side via an ingenious rescaling of parameters,
increasing $\lambda$ at fixed $\omega_D$ and $d$, thus decreasing $d/
\tilde\Delta$; however, the total number of levels $2 \omega_d/ d$ stays fixed
in the process, thus this way of reducing the effective level spacing, apart
from being (purposefully) unphysical, can only yield indirect and incomplete
information about the crossover, since it captures only the influence of the
levels closest to $\varepsilon_F$. Our method captures the crossover fully
without any such rescalings.

\begin{figure}
  \begin{center}
    \vspace{-3mm}
    \leavevmode
    \epsfxsize65mm
    \epsffile{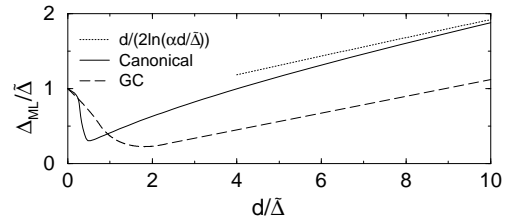}
    \vspace{-2mm}
  \end{center}
  \caption{The canonical (solid), g.c.\ (dashed)
    and analytical (dotted line) results for the parity parameter
    $\Delta_{\text{ML}}$ \protect\cite{Matveev-97}. }
  \label{fig:Delta-ML}
\end{figure}

\paragraph*{Matveev-Larkin's parity parameter.---}
ML \cite{Matveev-97} have introduced a parity parameter, defined to be the
difference between the ground state energy of an odd state and the mean energy
of the neighboring even states with one electron added and one removed:
$\Delta_{\text{ML}} = {\cal E}_1 - \frac12({\cal E}_0^{\text{add}} + {\cal
  E}_0^{\text{rem}})$.  Figure~\ref{fig:Delta-ML} shows the canonical and
g.c.\ results for $\Delta_{\text{ML}}$, and also the large-$d$ approximation
given by ML, $\Delta_{\text{ML}} = d/(2\log(\alpha d/\tilde\Delta))$, where the
constant $\alpha$ (needed because ML's analysis holds only with logarithmic
accuracy) was used as fitting parameter (with $\alpha=1.35$).  As for the
spectral gaps, the canonical and g.c. results are qualitatively similar,
though the latter of course misses the fluctuation-induced logarithmic
corrections for $d>d^C$.

In summary, the crossover from the bulk to the f.d.\ regime can be captured in
full using a fixed-$N$ projected BCS ansatz.  With increasing $d$, the pairing
correlations change from being strong and localized within $\tilde\Delta$ of
$\varepsilon_F$, to being mere weak, energetically delocalized
``fluctuations''; this causes the condensation energy to change quite
abruptly, at a characteristic spacing $d^C \propto \tilde\Delta$, from being
{\em extensive}\/ to {\em intensive}\/ (modulo small corrections).  Thus, the
qualitative difference between ``superconductivity'' for $d <d^C$, and
``fluctuations'' for $d > d^C$, is that for the former but not the latter,
adding {\em more}\/ particles gives a {\em different}\/ condensation energy;
for superconductivity, as Anderson put it, ``more is different''.

We would like to thank R. Fazio, G. Falci, A. Mastellone for sending us their
numerical data, and K. Likharev,
T. Poh\-jola, D. Ralph, A.\ Rosch, G. Sch\"on, and A.
Zaikin for discussions. This research was supported by ``SFB~195'' of the
Deutsche For\-schungs\-ge\-mein\-schaft and the German National Merit
Foundation.


\widetext

\begin{references}
\bibitem{likharev} The bulk and f.d.\ regimes differ, too, in the
  $\lambda$-dependence at fixed $d$ of $E_b$, which we found to be roughly
  $\sim e^{-2/\lambda}$ and $\sim \lambda$, respectively (as suggested to us
  by Likharev).
\bibitem{alternative} $\bar \Delta_b \equiv \lambda d \sum_j \langle
  c_{j+}^\dagger c_{j-}^\dagger c_{j-} c_{j+} \rangle \langle c_{j-} c_{j+}
  c_{j+}^\dagger c_{j-}^\dagger \rangle$, an alternative pairing parameter
  proposed in \cite{Braun-98}, turns out to be identically equal to $\Delta_b$
  for the ansatz $|b \rangle$.
\end{references}
\end{document}